\documentclass[onecolumn,usenatbib]{mn2e}
\usepackage{natbib}
\usepackage{amsmath}
\usepackage{graphicx}
\def\be{\begin{equation}}
\def\ee{\end{equation}}
\def\beq{\begin{eqnarray}}
\def\eeq{\end{eqnarray}}
\def\nn{\nonumber}
\begin{document}

\title{Modelling magnetically deformed neutron stars} 

\author[B. Haskell et al]
{B. Haskell$^1$, L. Samuelsson$^1$, K. Glampedakis$^{1,2}$, N. Andersson$^1$  \\
$^1$ School of Mathematics, University of Southampton, 
Southampton, SO17 1BJ, United Kingdom\\
$^2$ SISSA, via Beirut 2-4, 34014 Trieste, Italy}

\maketitle

\begin{abstract}
Rotating deformed neutron stars are important potential sources for groundbased gravitational-wave interferometers such as LIGO, GE0600 and VIRGO.
One mechanism that may lead to significant non-asymmetries is the internal magnetic field. 
It is well known that a magnetic star will not be spherical and, if the magnetic axis is not aligned with the spin axis, the deformation will lead 
to the emission of gravitational waves. The aim of this paper is to develop a formalism that would allow us to model 
magnetically deformed stars, using both realistic equations of state and field configurations. As a first step,
we consider a set of simplified model problems. Focusing on dipolar fields, we determine the internal  magnetic field  which is consistent with a given neutron star model. We then calculate the associated deformation. We conclude by discussing the relevance of our results for current 
gravitational-wave detectors and future prospects.
\end{abstract}

\maketitle

\section{Introduction}

It is well known that a non-axisymmetric deformation of a rotating neutron star will lead to a time varying quadrupole moment and 
could provide a good source of gravitational waves. In fact, rapidly rotating neutron stars are important potential 
sources of continuous gravitational waves for interferometric detectors such as LIGO, GEO600 and VIRGO as well as
the planned next generation interferometers which should be able to target them specifically.  There have even been suggestions 
of narrow-banding Advanced LIGO in order to target the Low Mass X-Ray Binaries (LMXBs) [see for example \citet{Brady}], 
as it is thought that gravitational waves may be playing a role in setting the spin equilibrium period of these systems \citep{bildsten,accrete} 

There are a number of mechanisms that may lead to a neutron star being deformed away from symmetry. 
First of all, the crust of a neutron star is elastic and can support ``mountains''. The size of the deformation that can be sustained depends on many factors, such as the equation of state and the evolutionary history of the crust. This problem has been studied by \cite{UCB} and \cite{haskell}. Another mechanism for producing asymmetries is an oscillation mode developing an instability, driven by gravitational radiation reaction, such as the r-mode instability proposed by \citet{rmode} (see \citet{review} for a relatively recent review). Deformations can also be caused by the magnetic field. Neutron stars are known to have significant magnetic fields and it is well known that a magnetic star will not remain spherical \citep{cf}. If the magnetic axis is not aligned with the rotation axis the deformation will not be axisymmetric and this will lead to gravitational wave emission. \citet{cutler2002} has suggested that a strong toroidal field could force a precessing neutron star to become unstable and  ``flip'' to become an orthogonal rotator. This would be an optimal configuration for gravitational-wave emission. \cite{heyl2000} has suggested that magnetised white dwarfs may be  interesting sources of gravitational waves. Several numerical studies have been directed at understanding the gravitational-wave emission of magnetically distorted neutron stars, for example \citet{BG1996}. Finally, \citet{melatos} have studied the closely related problem of magnetically confined accretion, which may lead to large deformations.

The aim of this paper is to investigate deformations due to the interior magnetic field in more detail. We want to develop a framework that 
would allow us to quantify the relevance of these asymmetries for any given stellar model and magnetic field configuration. As a starting point 
it makes sense to study some simple model problems. This will give us a better idea of the nature of the problem and what the key issues are. 
Moreover, it is a natural way to proceed given that the interior field structure is uncertain. The best current models [see, for example, 
\citet{spruit}]
tell us that the field will tend to have a mixed poloidal and toroidal nature. Any analysis should be able consider such generic 
configurations. 

It is, of course, the case that magnetically deformed stellar models have been studied previously and we can benefit greatly from the 
existing literature. It is particularly important to appreciate that the magnetic field configuration is constrained for any given stellar model.
In effect, the magnetic field must be solved for together with the fluid configuration. This statement is quite obvious, but it has not always been
appreciated in discussions of gravitational waves from magnetically deformed stars. Hence, we feel that it is appropriate to pay special attention to it here. This means that our paper has two main parts. The first part, which comprises Sections~2-4, is focused on the issue of the 
permissible field configuration. In many respects it is an adaptation of already existing results. It is nevertheless important to have this discussion, 
since it provides the main input for the deformation calculation. This calculation, the second part of the paper, is presented in Section~5.
The potential impact of our results on gravitational-wave observations is discussed in the concluding section. 
Throughout the paper, we consider only dipolar fields, but the developed formalism is general and can easily be extended to any magnetic field. 

\section{Magnetic fields in stellar interiors}

\citet{cf} where the first to realise that a star would not remain spherical in the presence of a strong magnetic field. They calculated the deformation of an incompressible star with a constant dipolar field by minimising the energy of the configuration. The case of a constant density star with an internal poloidal field matched to an external dipole was later considered by \citet{Fer} and \citet{goos}, by solving the Euler equations. In our analysis we shall use the latter approach.
However, before considering magnetic deformations it is important to understand which are the permissible field configurations, given the equation of state. We shall see that the range of permitted fields is quite restricted, a fact which is often overlooked when discussing magnetic deformations.
Let us, therefore, move on to describing the equilibrium configuration of a magnetic star. We will do this by assuming that the magnetic energy is small compared to the gravitational energy and that magnetic effects can be treated as a perturbation of a spherical, non magnetic, background. 
This should always be the case. 

The equations of hydrostatic equilibrium are:
\be
\frac{\boldsymbol{\nabla} p}{\rho}+\boldsymbol{\nabla}\Phi=\frac{(\boldsymbol{\nabla}\times \mathbf{B})\times \mathbf{B}}{4\pi\rho} = 
{\mathbf{L} \over 4 \pi \rho}
\label{hydro1}\ee
where $\mathbf{B}$ is the magnetic field, $p$ the pressure, $\rho$ the density and $\mathbf{L}$ defines the 
Lorentz force. The gravitational potential $\Phi$ obeys Poisson's equation
\be
\nabla^2\Phi=4\pi G\rho.
\label{poiss}\ee
The magnetic field must also obey, from Maxwell's equations,
\be
\boldsymbol{\nabla}\cdot \mathbf{B}=0
\label{div}\ee 
%and we shall (for simplicity) take a polytropic equation of state
%\be
%p=K\rho^{1+1/n}.
%\ee
Following \cite{Roxburgh} we take the curl of equation (\ref{hydro1}). This leaves us with an equation for the magnetic field
\be
\boldsymbol{\nabla}\times\left(\frac{\mathbf{B}\times(\boldsymbol{\nabla}\times \mathbf{B})}{\rho}\right)=0
\label{field}\ee
Since this equation contains the density it should be solved simultaneously with equation (\ref{hydro1}).
In other words,  the magnetic field structure is constrained by the density profile.

We now assume that the magnetic field only produces small deviations from a spherically symmetric background 
model (essentially, we assume that the ratio of magnetic to gravitational potential energy is small). 
This allows us to expand all our variables in the form
\be
\psi(r,\theta)=\psi_0(r)+\psi_1(r)P_{l}
\ee
where $P_{l}$ are the standard Legendre polynomials and $\psi_1$ is a small perturbation of $O(B^2)$. 
In the following we will concentrate on quadrupole ($l=2$) deformations, simply because they are optimal from the gravitational-wave 
emission point of view, However, the formalism  applies to the general problem, in which case the perturbation
is given by a sum of Legendre polynomials.

We can first of all solve the structure equations in the absence of a magnetic field and obtain the background model. The result is then fed into equation (\ref{field}) to determine $\mathbf{B}$, which we will need to solve equation (\ref{hydro1}) and (\ref{poiss}) to first perturbative 
order for the quantities $\rho,p$ and $\Phi$.
Restricting ourselves to axisymmetry, the $\phi$ component of the magnetic force must be zero, as there is nothing to balance it in equation (\ref{hydro1}). Hence
\be
\left[\mathbf{B}\times(\boldsymbol{\nabla}\times \mathbf{B})\right]_{\phi}=0
\label{zero}
\ee
With this understanding, let us examine some general magnetic field solutions for various background models.
If one splits the magnetic field into two components, a poloidal one $\mathbf{B}_p=(B_r,B_{\theta},0)$ and a toroidal one $\mathbf{B}_t=(0,0,B_{\phi})$, and introduces a stream function $S(r,\theta)$ such that
\beq
B_r&=&\frac{1}{r^2\sin{\theta}}\frac{\partial S}{\partial\theta}\nn\\
B_\theta&=&-\frac{1}{r\sin{\theta}}\frac{\partial S}{\partial r}
\eeq
then equations (\ref{div}) and (\ref{zero}) reduce to \citep{Roxburgh}:
\be
\mathbf{B}_p\cdot\boldsymbol{\nabla}(r\sin{\theta}\mathbf{B}_t)=0
\ee
which gives
\be
B_\phi=\frac{\beta(S)}{r\sin{\theta}}
\ee
Where $\beta$ is some function of the stream function $S$. This means that the toroidal part of the field is a function of the poloidal part.
Equation (\ref{field}) then takes the form

\beq
&&\frac{\partial}{\partial r}\left\{\frac{1}{\rho r\sin{\theta}}\frac{\partial S}{\partial\theta}\left[\frac{1}{r\sin{\theta}}\frac{\partial^2S}{\partial r^2}+\frac{1}{r^3}\frac{\partial}{\partial\theta}\left(\frac{1}{\sin{\theta}}\frac{\partial S}{\partial\theta}\right)\right]+\frac{\beta}{\rho r^2\sin^2{\theta}}\frac{\partial\beta}{\partial\theta}\right\}-\frac{\partial}{\partial\theta}\left\{\frac{1}{\rho r\sin{\theta}}\frac{\partial S}{\partial r}\left[\frac{1}{r\sin{\theta}}\frac{\partial^2S}{\partial r^2}\right.\right.\nn\\
&&\left.\left.+\frac{1}{r^3}\frac{\partial}{\partial\theta}\left(\frac{1}{\sin{\theta}}\frac{\partial S}{\partial\theta}\right)\right]+\frac{\beta}{\rho r^2\sin^2{\theta}}\frac{\partial\beta}{\partial r}\right\}=0
\label{stream}
\eeq
In the following we shall focus on dipole solutions. We can then take a stream function of form
\be
S(r,\theta)=A(r)\sin^2{\theta}
\label{dipo}\ee
or directly look for solutions of the form:
\be
\mathbf{B}=\mathbf{\hat{r}}\{W(r)\cos{\theta}\}+\boldsymbol{\hat{\theta}}\{X(r)\sin{\theta}\}+\boldsymbol{\hat{\phi}}\{iZ(r)\sin{\theta}\}
\label{dipole}\ee
where the phase of the $\boldsymbol{\hat{\phi}}$ component
is chosen in such a way as to produce real valued equations in the following.  
For this kind of field, equation (\ref{div}) gives
\be
rW{'}(r)+2\left[W(r)+X(r)\right]
\label{divspec}
\ee
where the prime indicates differentiation with respect to $r$.
We now need to solve equation (\ref{field}) [or equivalently equation (\ref{stream})], which depends on the density profile $\rho(r)$.
It is thus necessary to prescribe an equation of state for the stellar matter.

%%%%%%%%%%%%%%%%%%%%%%%%%%%%%%%%%%%
\section{Uniform density stars}
%%%%%%%%%%%%%%%%%%%%%%%%%%%%%%%%%%%

In order to begin investigating the problem and understand how restricted our choice of magnetic field configurations may be, it is useful to consider a constant density star.
Insert $\rho$=constant into (\ref{field}) to get
\beq
&W\left(rZ\right){'}+2ZX=0\label{lars1}\\
&W\left(r^2Z{''}-2Z\right)+2Z\left(rX{'}-X\right)=0\label{lars2}\\
&W\left[r^2X{''}-4\left(W+X\right)\right]+2Z\left(rZ{'}-Z\right)=0\label{lars3}\eeq

\subsection{Poloidal fields}

Let us first restrict our attention to purely poloidal fields.
In this case we can combine equation (\ref{lars3}) with equation (\ref{div}) to get
\be
r^2X{'''}+4rX{''}-4X{'}=0
\ee
which has the solution
\be
X=D+Cr^2
\label{prima1}\ee
with $C$ and $D$ constants and where the $\propto\, r^{-3}$ solution has been excluded to ensure regularity at the centre. This gives, for $W$,
\be
W=-D-\frac{1}{2}Cr^2
\label{prima2}\ee
Note that for $C=0$ this corresponds to a uniform dipole field.
The same solution can be obtained by solving for the stream function. It is then 
sufficient to take $S(r,\theta)$ of the form in equation (\ref{dipo}) and take $\beta=0$ so that the magnetic field reduces to
\be
\mathbf{B}=\left(\frac{2A\cos{\theta}}{r^2},\frac{-A{'}\sin{\theta}}{r},0\right)
\ee
We can now solve equation (\ref{stream}) with the boundary conditions that the field must remain finite at the centre. i.e.
\be
\frac{A}{r^2},\frac{A{'}}{r}\;\;\mbox{finite at $r=0$}
\label{finite}\ee
while at the surface the field must be continuous with an external curl free dipole field, so that we have
\be
\frac{A}{r}+A{'}=0 \ , \qquad \mbox{at } r=R .
\ee
The solution to this problem was given by \cite{Fer}, who considered a constant density star with a current density of the form
\be
J=\frac{B}{R^2}r^3\sin^3\theta
\ee
where $B$ is a constant parameter. In this case equation (\ref{stream}) reduces to
\be
\nabla^2 S = \frac{B}{R^2}r^2\sin^2\theta
\ee
which yields for the stream function:
\be
S=B\frac{r^2}{2 R^2}\left(\frac{r^2}{5}-\frac{R^2}{3}\right)\sin^2{\theta}
\ee
The magnetic field is then
\beq
B_r&=& - B\left[\frac{1}{3}-\frac{1}{5}\left(\frac{r}{R}\right)^2\right]\cos{\theta}\nn\\
B_\theta&=&B\left[\frac{1}{3}-\frac{2}{5}\left(\frac{r}{R}\right)^2\right]\sin{\theta}
\label{Fer}\eeq
which corresponds to taking 
\be
C=-\frac{2}{5}\frac{B}{R^2}\;\;\;\mbox{and}\;\;\;D=\frac{B}{3}
\ee
in the previous solution (\ref{prima1})-(\ref{prima2}). 

In Section~5 we will work out how this field
deforms the shape of the star. As we will consider several other field configurations, and it would be useful to be able to compare the results, 
it is worth thinking about how such a comparison would be carried out. After all, each model field will be naturally described 
by some set of parameters which may not be easily translatable from model to model. Intuitively, one would expect the 
deformation of the star to depend on the ratio of magnetic to gravitational potential energy [see for example \citet{cutler2002}]. 
Hence it would make sense to consider the magnetic energy as a useful measure. Introducing the volume averaged magnetic field energy $\bar{B}$, 
one can show that for the field in (\ref{Fer}) we have $\bar{B}=0.1 B$.

%%%%%%%%%%%%%%%%%%%%%%%

\subsection{Toroidal field}

For a purely toroidal field, on the other hand, the equations reduce to
\be 
rZ^{'}=Z
\ee
which gives
\be
Z=-icr
\label{soltoro1}\ee
This corresponds to a uniform current distribution inside the star, i.e. $|{\boldsymbol{\nabla}}\times {\mathbf{B}}|=2c=$constant, pointing along the $z$-axis .

\subsection{Mixed poloidal/toroidal field}

Let us return to the general case. If we formally solve equation (\ref{divspec}) for $X$, then equation (\ref{lars1}) becomes
\be
r^2\left[W \left(\frac{Z}{r}\right)^\prime-\left(\frac{Z}{r}\right)W{'}\right]=r^2W^2\left(\frac{Z}{rW}\right)^\prime=0
\ee
Hence, unless $W=0$ or $Z=0$ we have
\be
Z=arW
\label{dipende}\ee
where $a$ is a constant.
Note that equation (\ref{lars2}) is implied by (\ref{lars1}) by the use of (\ref{divspec}); thus the substitutions 
\be
Z=arW\;\;\;\;\mbox{and}\;\;\;\;X=-W-\frac{r}{2}W^{'}
\ee
solve equations (\ref{lars1}) and (\ref{lars2}). The last equation, (\ref{lars3}), then becomes
\be
\frac{1}{2}rW\left(r^2W^{'''}+4rW{''}-r(1+a^2r^2)W^{'}\right)=0
\ee
Thus we find that the solution can be written 
\be
W=\frac{1}{r^3}[C_1(2ar-1)e^{2ar}+C_2(2ar+1)e^{-2ar}]+C_3
\ee
leading to
\be
X=-\frac{1}{2r^3}[C_1(4a^2r^2-2ar+1)e^{2ar}-C_2(4a^2r^2+2ar+1)e^{-2ar}]-C_3
\ee
where $C_i$ are constants. In order to ensure regularity at the centre we must take $C_1=C_2$, and the central values of the fields (which we shall denote with a subscript $c$) then become
\be
W_c=-X_c=\frac{16}{3}C_1a^3+C_3\;\;\;\;\mbox{and}\;\;\;\;Z_c=0\ee
As we can see, the parameter $a$ is just a length scale and can be absorbed in the other constants if we define a new dimensionless radial coordinate $x=2ar$. This allows us to redefine the constant
\be
C_1 \rightarrow C_1/(2a)^3
\ee
The free parameters in the regular solution now correspond to the new $C_1$ and $C_3$ and are only two (if we exclude the trivial choice of scale or units given by $a$). Explicitly we have
\beq
W&=&\frac{2C_1\cosh x}{x^2}-\frac{2C_1\sinh x}{x^3}+C_3\nn\\
X&=&\frac{C_1\cosh x}{x^2}-\frac{C_1(x^2+1)\sinh x }{x^3}-C_3\nn\\
Z&=&\frac{C_1\cosh x }{x}-\frac{C_1\sinh x }{x^2}+\frac{1}{2}x C_3
\eeq
We can further interpret the parameters by noting that the central values of the fields are $W_c=-X_c=\frac{2}{3}C_1+C_3$ ($Z_c=0$ still). We can thus use $C_1$ as an overall scale for the field and define $\hat{W}=W/C_1$ and likewise for the other variables. Thus, using $C_3/C_1=\hat{W}_c-2/3$, we obtain
\beq
\hat{W}&=&\frac{2\cosh x }{x^2}-\frac{2\sinh x }{x^3}+\hat{W}_c-2/3\nn\\
\hat{X}&=&\frac{\cosh x }{x^2}-\frac{(x^2+1)\sinh x }{x^3}-\hat{W}_c+2/3\nn\\
\hat{Z}&=&\frac{\cosh x }{x}-\frac{\sinh x }{x^2}+\frac{1}{2}x(\hat{W}_c-2/3)
\eeq
We now have only one non-trivial parameter, $\hat{W}_c$, which controls the ratio of poloidal to toroidal field, while the other parameters $a$ and $C_1$ have been absorbed into the definition of our variables. They can simply be interpreted as scales of the problem (specifically a length scale and the scale of the magnetic field). 

\subsection{Boundary conditions}

Let us now focus on the boundary conditions that the magnetic field needs to satisfy at the surface of the star (at the centre it is sufficient to impose regularity). The solenoidal nature of the field requires continuity of the radial component
\be
\left<B^r\right>=0
\ee
where $\left<B^r\right>=B^r_\mathrm{ext}-B^r_\mathrm{int}$ 
indicates the discontinuity between the external and internal parts of the field at the surface.
Furthermore, we shall require continuity across the boundary of the traction vector $t^i=T^{ij}\hat{n}_j$, i.e. the projection of the total stress tensor along the normal unit vector\footnote{While the main part of our discussion uses standard spherical coordinates, the discussion of the boundary conditions becomes clearer if we use a coordinate basis. This means that the Einstein convention of summation over repeated indices applies.}. 
This is equivalent to requiring local force balance per unit area. Our condition is hence
\be
\langle t^i \rangle=0
\ee
The stress tensor is the sum of a fluid piece and a magnetic piece
\be
T^{ij}=-p\delta^{ij}+\frac{1}{4\pi}\left(B^iB^j-\frac{1}{2}B^2\delta^{ij}\right)
\ee
Let us consider the projection along the normal to the surface. The normal vector to the perturbed surface will have the form
\be
\hat{n}_S=\hat{r}+\hat{n}_S^1
\ee
where $\hat{n}_S^1$ indicates the correction of $O(B^2)$.
Projecting $T^{ij}$ along this vector, we obtain
\be
-p^0\delta^{ij}\hat{n}_{S}^1-\delta p\,\,\delta^{ir}+\frac{1}{4\pi}\left(B^iB^r-\frac{1}{2}B^2\delta^{ir}\right)
\ee
Where $p^0$ is the background pressure and $\delta p$ the first order perturbation. 
Note that the magnetic term is already $O(B^2)$ and one can take $\hat{n}_S=\hat{r}$ for this term.
At the surface the background pressure $p^0$ vanishes, so we must impose continuity for
\beq
t^r&=&-\delta p+\frac{1}{4\pi}[(B^r)^2-\frac{1}{2}B^2]\\
t^\theta&=&\frac{1}{4\pi}B^rB^\theta\\
t^\phi&=&\frac{1}{4\pi}B^rB^\phi
\label{boundary}\eeq
As we already have $\langle B^r \rangle=0$,  the $\theta$ and $\phi$ components of equations (\ref{boundary}) demand that 
$\langle B^{\theta}\rangle=\langle B^{\phi}\rangle=0$, which then leads to $\langle B^2\rangle=0$. As $p=0$ in the exterior, it must be that  $\delta p=0$ at the surface (all quantities are now $O(B^2)$ so we can consider the surface of the unperturbed configuration). 
The conclusion is thus that all components of the magnetic field must be continuous across the interface, i.e. that we have no surface currents. 

We should stress that we are in principle allowed to  introduce surface currents at this point. 
This would lead to discontinuities in the $\theta$ and $\phi$ components of the field. However, we feel that
unless there are physical arguments dictating the nature of such currents there is no reason to introduce them.
If we allow for the presence of an arbitrary current at the surface we have too much speculative freedom and it is not clear to what extent various models make sense. Hence, we find it more natural to restrict ourselves to models where the above  traction conditions apply.

Note that the traction conditions  are automatically satisfied if we have a purely toroidal field. 
The external magnetic field, in fact, then solves
\be
{\boldsymbol{\nabla}}\cdot {\mathbf{B}}={\boldsymbol{\nabla}}\times {\mathbf{B}}=0
\ee
and the assumption of axisymmetry forces a vanishing toroidal component $B^\phi=0$.
In the case of a purely poloidal field it is sufficient to match with an external curl-free dipole. 
In the case of a mixed poloidal/toroidal field, however, one must have that at the surface
\be
Z(R)=0
\ee
and as we have found that $Z=arW$, cf. (\ref{dipende}), it must also be the case (provided that $a\neq 0$) that $W=0$. This means that, under axisymmetry, a mixed poloidal and toroidal dipolar internal field and a general multipolar external field are forced to obey $B^r_\mathrm{int}=B^r_\mathrm{ext}=0$ and $B^{\phi}_\mathrm{int}=B^{\phi}_\mathrm{ext}=0$ at the surface.
It is thus immediately clear that we cannot match this kind of internal field with an external dipole field.
One also has a condition on $B^{\theta}$, as the exterior solution for a field which is regular at infinity is of the form
\be
\mathbf{B}_\mathrm{ext}=\sum_{l\geq m}\mathbf{\hat{r}}[-(l+1)\frac{A_l}{r^{l+2}}Y_l^m]+\boldsymbol{\hat{\theta}}[\frac{A_l}{r^{l+2}}\partial_\theta Y_l^m]+\boldsymbol{\hat{\phi}}[\frac{i m A_l}{r^{l+2}}\sin\theta Y_l^m]
\ee
It is clear that if the $\mathbf{\hat{r}}$ component must vanish at the surface, so must the ${\hat{\boldsymbol{\theta}}}$ component.
 We thus have that all components of the mixed magnetic field must vanish at the surface.
 As we shall see in the following, this is true also in the case of a polytropic equation of state. In conclusion one can, therefore, match an interior mixed poloidal/toroidal dipole field only to a vanishing external field.
It is also clear that the model does not allow for a purely toroidal field, since (\ref{soltoro1}) is only compatible with the 
surface condition if $c=0$. 
It is interesting to note how, as a consequence of the boundary conditions, we are restricted to a very limited class of magnetic fields: notably poloidal or mixed poloidal/toroidal fields that vanish outside the star.
One may wonder if this restriction derives from having taken a somewhat pathological and simplistic model, the constant density one. To establish that this is not the case, let us move on to a more realistic case.

\section{Polytropic models}

Having understood the restrictions of the uniform density case, let us now allow for a non-uniform density distribution. It is natural 
to consider an $n=1$ polytrope, as this model has many  features of a realistic neutron star model. It also permits an analytic treatment.
The allowed magnetic field must still satisfy equation (\ref{field}), but now the role of $\rho$ is no longer trivial as we are considering a density profile
\be
\rho(r)=\rho_c\frac{R}{r\pi}\sin\left({\frac{r\pi}{R}}\right)
\ee

\subsection{Poloidal field}

Following \cite{Monaghan1966}, we solve equation (\ref{stream}), imposing regularity at the centre and matching to an external dipole. The solution for the stream function is then
\be
S=-\frac{2}{3} \frac{\sin^2 {\theta}}{rR^2\pi} \left[ -{r}^{3}{\pi }^{3}+3R\,(2R^2-\pi^2r^2)\sin
 \left( {\frac {r\pi }{R}} \right) -6\,{R}^{2}r\pi \,\cos \left( {
\frac {r\pi }{R}} \right)  \right]
\ee
which leads to a field of form
\beq
B_r&=&\frac{B_s \cos \theta }{\pi R^3}  \left[ {r}^{3}{\pi }^{3}-3R\,(2R^2-\pi^2r^2)\sin \left( {\frac {r\pi }{R}} \right) +6\,{R}^{2}r\pi \,\cos \left( {\frac {r\pi }
{R}} \right)  \right]\frac{1}{\left( {\pi }^{2}-6
 \right)}\\
B_\theta&=&-\frac{B_s}{2}\frac{\sin \theta}{\pi R^3}  \left[ 2\,{r}^{3}{\pi }^{3}+3R(2R^2-\pi^2r^2)\,\sin\left( {\frac {r\pi }{R}} \right)-3\pi r(2R^2-\pi^2r^2)\,\cos \left( {
\frac {r\pi }{R}} \right) \right]\frac{1}{\left( {\pi }^{2}-6
 \right)}
\label{n1}\eeq
where $B_s$ represents the conventional dipole field strength at the surface.

\subsection{Mixed toroidal and poloidal fields}

Let us now assume a mixed poloidal and toroidal configuration. We will once again look for dipolar solutions and take a stream function of the form
$S(r,\theta)=A(r)\sin^2{\theta}$. Furthermore, following \citet{Roxburgh} we shall define $\beta$ to be 
\be
\beta=\frac{\lambda}{R} S
\ee
This is obviously a particular choice. However, it is natural since it leads to a separable equation
which means that the problem can be treated more or less analytically. 
By varying the parameter $\lambda$, which  describes the relative strength of the toroidal part of the field, one can hope to 
get some insight into the how the two field components interact in the mixed problem.

The magnetic field thus takes the form 
\be
\mathbf{B}=\left(\frac{2A\cos{\theta}}{r^2},\frac{-A{'}\sin\theta}{r},\frac{\lambda A\sin\theta}{rR}\right)
\label{genfield}
\ee
Then
 equation (\ref{stream}) can be written
\be
A\frac{d}{dr}\left[\frac{1}{\rho r^2}\left(\frac{2A}{r^2}-\frac{d^2A}{dr^2}\right)-\frac{\lambda^2A}{\rho R^2r^2}\right]=0
\label{st2}\ee
As before, we want to find a solution such that the field remains finite at the centre, cf. (\ref{finite}), 
and that all the components of the magnetic field be continuous at the surface.
 As the toroidal field must vanish at the surface this forces the condition
\be
A=0\;\;\mbox{at}\;\; r=R
\ee
It must also be the case that
\be
A{'}=0
\label{b2}
\ee
which derives from the condition that all components of the field must vanish at the surface.
We can thus only consider fields that vanish outside the star.

Solving the above equations for the density profile of an $n=1$ polytrope, we find that the 
stream function takes the form
\beq
A(r)&=&- \left[ \sin \left( {\frac {r\pi }{R}} \right) {r}^{2}{\lambda}^{3}\cos 
\lambda-\sin \left( {\frac {r\pi }{R}} \right) {
r}^{2} {\lambda}^{2}\sin \lambda-\lambda\,\sin \left( {
\frac {r\pi }{R}} \right) {r}^{2}\cos \lambda {\pi }^{2
}-2\,\lambda\,R\pi \,\cos \left( {\frac {r\lambda}{R}} \right) r\right.\nn\\
&&\left.-2\,\lambda\,R\pi \,r\cos \left( {\frac {r\pi }{R}} \right) +2\,\lambda\,\sin \left( {\frac {r\pi }{R}} \right) {R
}^{2}\cos  \lambda +\sin \left( {\frac {r\pi }{R}}
 \right) {r}^{2}\sin \lambda  {\pi }^{2}+2\,R\pi \,r\cos
 \left( {\frac {r\pi }{R}} \right) \sin  \lambda \right.\nn\\
&&\left.-2\,
\sin \left( {\frac {r\pi }{R}} \right) {R}^{2}\sin \lambda
 +2\,{R}^{2}\pi \,\sin \left( {\frac {r\lambda}{R}} \right) 
 \right]\frac{B_k{R}^{2}}{r F(\lambda)}
\label{tutto1}\eeq
where $B_k$  parametrises the strength of the magnetic field. 
We have defined
\be
F(\lambda)=\left[\lambda\cos\lambda-\sin\lambda\right](\lambda^2-\pi^2)^2
\ee
Imposing the boundary conditions we find an eigenvalue
relation for $\lambda$. Permissible fields must be such that 
\be
\tan\lambda=\frac{\lambda(\pi^2-\lambda^2)}{\pi^2-3\lambda^2}
\label{eig}\ee
By taking higher values of $\lambda$ we can increase the relative strength of the toroidal part of the field compared to the poloidal part.
Solving the trancendental equation numerically we find that  
the first three eigenvalues are
\beq
\lambda_1&=&7.420\nn\\
\lambda_2&=&10.706\nn\\
\lambda_3&=&13.917
\eeq
These results agree with the values given in \cite{Roxburgh} with an accuracy of $\approx 0.1\%$. The key point is that the toroidal part of the field 
is not freely specifiable in this model. If we prescribe the strength of the poloidal field component, then we can find solutions with increasing 
toroidal fields, but these are discrete. 

\subsection{Purely toroidal fields}

In the case of a purely toroidal field equation (\ref{field}) takes the form 
\be
\frac{\partial}{\partial r}\left(\frac{B_\phi}{\rho r\sin{\theta}}\right)\frac{\partial}{\partial\theta}(B_\phi r\sin{\theta})-\frac{\partial}{\partial\theta}\left(\frac{B_\phi}{\rho r\sin{\theta}}\right)\frac{\partial}{\partial r}(B_\phi r\sin{\theta})
\ee
The boundary conditions we have to impose are that the field vanish both at the centre  and at the surface of the star, as before.
We can take a solution of the simple form 
\be
B_{\phi}=\frac{B}{\pi}\sin\left(\frac{r\pi}{R}\right)\sin{\theta}
\label{toro1}
\ee
which will satisfy the boundary conditions since $\rho=0$ at the surface, 
as is the case for the polytropic model we are considering. Note that, in contrast to the uniform density problem, the polytropic equation of state
allows for a purely toroidal field. 

\subsection{Field confined to the neutron star ``crust''}

It is often assumed that, if the core of a neutron star is a type I superconductor the magnetic field will be expelled from the core and confined to the crust, see for example \citet{BG1996}. To discuss this situation one should in principle consider the full equations of hydrostatic equilibrium, including the elastic terms. However, in order to investigate this problem, we will consider the case of a fluid with a magnetic field confined to a region close to the surface. This model would be relevant provided that the deformed shape represents the relaxed configuration of the crust. Then 
there is no strain, and thus no elastic terms in the equilibrium equations. We shall consider the same mixed 
field as in the previous section, i.e. of the form (\ref{genfield}). We shall, however consider the field to vanish inside a certain radius $r_b$ (which can be considered to be the base of the crust). The boundary conditions for the third order differential equation (\ref{st2}) at the surface thus remain
$A(R)=A\,{'}(R)=0$
together with a condition which comes from imposing continuity of the $B_r$ component of the magnetic field, i.e. imposing
\be
A(r_b)=0
\label{crust2}\ee
We  again get an eigenvalue problem for the parameter $\lambda$, allowing us to calculate the permitted ratios of toroidal to poloidal field strength.
It should be noticed that we are not imposing continuity of the tractions at the inner boundary. In fact, there is a  discontinuity in the $B^{\theta}$ component of the field at $r_b$, which will lead to currents at the crust/core interface. In order to simplify the calculation we also take the core to be unperturbed, and simply impose continuity of the perturbation in the gravitational potential $\delta\Phi$ and of its derivative $\delta\Phi^\prime$.

Before moving on, it is worth remarking that this simple model can only be seen as a rough representation of the problem with a superconducting core. 
There are a number of issues that one ought to worry about, and which we are not considering here. For example, the complete flux expulsion from a Type~I superconducting core may be a severe oversimplification, Is it really to be expected that the magnetic flux is expelled from the entire core rather than
(say) bunched up into macroscopic regions? Moreover, if the core forms a type~II superconductor (as is usually expected) then the magnetic flux is 
carried by fluxtubes. This problem is significantly different from our idealized model, and requires a separate analysis. 

\section{Magnetic deformations}

So far we have discussed how the requirement that the magnetic 
field to be consistent with the fluid configuration greatly restricts the model parameter space. In particular, 
it does not allow us to freely specify the toroidal component of the field in the mixed case. 
Also, for the models that we have considered,  one can only 
consider fields which vanish in the exterior. This is clearly not physical, as one would expect the exterior field to be prevalently dipolar far from the star. This problem could easily be fixed by allowing surface currents (which one may expect as there are strong electric fields at the stellar surface). Nevertheless, given that we do not have a physical model for such currents we will not introduce them here.

Having determined some magnetic field configurations which are consistent with a chosen stellar interior, we can turn our attention to solving equations (\ref{hydro1}) in order to obtain the new equilibrium shape of the star. Inspired by the work of \citet{saio} and our own recent work on crustal deformations \citep{haskell}, we shall define a new variable
\be
x(r,\theta)=r\left[1+\varepsilon(r)P_l(\theta)\right]\label{newvar}
\ee
where $r$ is the standard radial variable and $P_l$ is a Legendre polynomial representing the deformation. 
Note that we are considering only one polynomial here, but the formalism could  easily be extended to a sum of Legendre polynomials. 
The perturbed surface thus takes the form
\be
x_s(R,\theta)=R\left[1+\varepsilon(R)P_l(\theta)\right]
\ee
where $R$ is the radius of the unperturbed (spherical) star.
We shall also assume that the pressure, density and gravitational potential take the form
\be
\psi(r,\theta)=\psi(r)+\delta\psi^l(r)P_{l}
\ee
where $\psi$ is the background quantity and $\delta\psi^l$ is a small perturbation of $O(B^2)$. From now on we shall, unless there is a risk of confusion, write $\delta\psi$ instead of $\delta\psi^l$.
Equations (\ref{hydro1}) thus take the form
\beq
&&\left(\frac{d\delta p}{dr}+\rho\frac{d\delta\Phi}{dr}+\delta\rho\frac{d\Phi}{dr}\right)P_l=\frac{\left[({\boldsymbol{\nabla}}\times {\mathbf{B}})\times {\mathbf{B}}\right]_r}{4\pi}\\
&&\left(\delta p+\rho\delta\Phi\right)\nabla P_l=\frac{\left[({\boldsymbol{\nabla}}\times{\mathbf{B}})\times {\mathbf{B}}\right]_\theta}{4\pi}
\label{hydro2}\eeq
and  must be solved together with the perturbed Poisson equation
\be
\frac{d^2\delta\Phi}{dr^2}+\frac{2}{r}\frac{d\delta\Phi}{dr}-\frac{6}{r^2}\delta\Phi=4\pi\delta\rho
\label{pertp}\ee
Let us first of all tackle the case of an incompressible star.

%%%%%%%%%%%%%%%%%%%%%%%%%%%%%%%%%%%%%%%%%%%%%%%%%%%%%%%%%%
\subsection{Deformations of incompressible stars}
%%%%%%%%%%%%%%%%%%%%%%%%%%%%%%%%%%%%%%%%%%%%%%%%%%%%%%%%%%%%%%

In the case of a uniform density star we consider the field in equation (\ref{Fer}). This gives us for the Lorentz force
\beq
L_r &=& { 2 \over 3} \left({B \over R^2} \right)^2\left(\frac{R^2r}{3}-\frac{2r^3}{5}\right)\left(1-P_2\right)\\
L_\theta&=&-\left({B \over R^2} \right)^2\left(\frac{R^2r}{6}-\frac{r^3}{10}\right)\left(\frac{1}{3}\frac{dP_2}{d\theta}\right)
\eeq
as well as, obviously, $L_\phi=0$.
In the case of an incompressible star, there can be no $l=0$ deformation, as this would not conserve the volume and therefore not conserve the mass of the star. We shall thus only consider the case of $l=2$. For this case $\delta\Phi$ inside the star takes the form 
\be
\delta\Phi=-\frac{4}{5}\pi G\rho\varepsilon(R)r^2
\label{deltaphi}
\ee
Inserting this solution into the $\hat{\boldsymbol{\theta}}$ component of (\ref{hydro2}) and evaluating at the surface gives
\be
\delta p(R)=\frac{4}{5}\pi G\rho\varepsilon(R)R^2-\frac{B^2}{90\pi}\label{pofR}
\ee
If we now remember that
\be
\delta p(r)=\delta p(x)-\varepsilon(r)r\frac{dp}{dr}(r)
\ee
we have at the surface
\be
\delta p(R)=0-\varepsilon(R)R\frac{dp}{dr}(R)
\label{psurf}
\ee
By using the background pressure 
\be
p=2\pi G\rho^2\frac{1}{3}(R^2-r^2),
\ee
equation (\ref{pofR}) gives us
\be
\varepsilon(R)=-\frac{1}{48}\frac{B^2}{\pi^2G\rho^2 R^2}
\ee
which agrees with the result of \cite{goos}.

In order to assess the relevance of this deformation for gravitational-wave emission we
calculate the ellipticity, which is defined as
\be
\epsilon=\frac{I_{zz}-I_{xx}}{I_0}
\ee
Here $I_0$ is the moment of inertia of the spherical star, while $I_{jk}$ is the inertia tensor
\be
I_{jk}=\int_V\rho(r)(r^2\delta_{jk}-x_jx_k)dV.
\ee
 For a constant density star this leads to
\beq
\epsilon&=&-\frac{3}{2}\varepsilon=\frac{1}{18}\frac{B^2R^4}{GM^2}\approx 10^{-12}\,\left(\frac{R}{10\mbox{ km}}\right)^4\,\left(\frac{M}{1.4 M_{\odot}}\right)^{-2}\,\left(\frac{\bar{B}}{10^{12}G}\right)^2
\eeq
where $\bar{B}$ is the energy averaged magnetic field.
The ellipticity is positive, so the star is oblate, as expected.
As a sanity check of this result we can compare to the estimate used by \citet{cutler2002}. For a field of $10^{15}$~G he assumes that 
$\epsilon\approx 1.6 \times 10^{-6}$. Our more detailed calculation has led to a result that is almost a factor of two smaller
than this. Nevertheless, the two estimates are reasonably close.

%%%%%%%%%%%%%%%%%%%%%%%%%%%%%%%%%%%%%%%%%%%%%%%%%%%%%%%%%%%%%
\subsection{n=1 polytrope with a poloidal field}
%%%%%%%%%%%%%%%%%%%%%%%%%%%%%%%%%%%%%%%%%%%%%%%%%%%%%%%%%%%%%%%%%%

Let us now consider the case of a star with an $n=1$ polytropic equation of state. The magnetic field appropriate for this case is that in equation (\ref{n1}), for which the Lorentz force has non-trivial $l=2$ components\footnote{In the general case there will be both a radial and a quadrupole deformation. Since we are primarily interested in the gravitational-wave aspects we ignore the $l=0$ deformation in the discussion. This contribution would be important if one wanted to (say) study the oscillations of deformed magnetic stars. The required calculation is provided in Appendix~B.}  
\beq
L_r\!\!&=&\!\!-\frac{1}{2} \left[ 2\,{r}^{3}{\pi }^{3}+\left( 6R^3-3R{r}^{2}{\pi }^{2}\right)\sin \left( {\frac {r\pi }{R}}
 \right)+\left(3r^3\pi^3-6{R}^{2}r\pi\right)\cos \left( {\frac {r\pi }{R}} \right) \right]\frac{ {\pi}^{2}{B_s}^{2}}{ {r}\left( {\pi }^{2}-6 \right){R}} \sin \left( {\frac {r\pi }{R}}\right)  P_2(\theta)\\
&&\nn\\
L_\theta\!\!&=&\!\!-\frac{1}{2} {\pi }^{2} \left[ {r}^{3}{\pi }^{3}-\left( 6R^3-3R{r}^{2}{\pi }^{2}\right)\sin \left( {\frac {r\pi}{R}} \right) +6 {R}^{2}r\pi \cos \left( {\frac {r\pi }{R}}
 \right)  \right]\frac{{B_s}^{2}}{ {r}\left( {\pi }^{2}-6 \right){R}}\sin \left( {\frac {r\pi }{R}} \right) \frac{dP_2}{d\theta}
\label{lorentzn1}\eeq
Having worked out the Lorentz force we can solve the perturbed hydrostatic equilibrium equations (\ref{hydro2}) and the Poisson equation (\ref{pertp}), with the condition that $\delta\Phi$ and $\delta\Phi'$ be regular at the centre and match the exterior solution at the surface.
From equation (\ref{hydro2}) we then obtain
\be
\delta\rho=-\Big[ {r}^{3}{\pi }^{6}+(3Rr^2\pi^5-6R^3\pi^3)\sin \left( {\frac {\pi \,r}{R}} \right)+6\,{\pi }^{4}{R}^{2}r
\cos \left( {\frac {\pi \,r}{R}} \right)+2{\rho_c}{
\delta\Phi} \left( r \right) \left(\frac{R}{B_s}\right)^{2}r(\pi^2-6)^2\Big]\frac{B_s^2\pi}{8r( {\pi^2 }-6) ^{2}{G}{{\rho_c}}{R}^{4}}
\ee
Inserting this into the right hand side of equation (\ref{pertp}) we can solve for $\delta\Phi$
\be
\delta\Phi=\left[ {r}^{5}{\pi }^{3}\cos \left( {\frac {\pi \,r}{R}}
 \right) -2\,{r}^{5}{\pi }^{3}-(3r^4\pi^2R-36R^5+12\,{\pi }^{2}{r}^{2}{R}^{3})\sin
 \left( {\frac {\pi \,r}{R}} \right)-36\,\pi \,r{R}^{4}\cos \left( {\frac {\pi \,r}{R}} \right)\right] \frac{{\pi }^{3}{B_s}^{2}}{4\left( {\pi }^{2}-6 \right) ^{2
}{r}^{3}{R}^{2}{{\rho_c}}}
\ee
We can then evaluate the distortion at the surface:
\be
\varepsilon=-\frac{\delta\rho}{R}\frac{d\rho}{dr} = {\frac {{\pi }^{5} \left( {\pi }^{2}-24 \right) {B_s}^{2}}{16{R}^{2}
{\rho_c}^{2}G \left( {\pi }^{2}-6 \right)^2 }}
\ee

Finally we obtain the ellipticity   
\be
\epsilon=-{\frac {{\pi }^{5}R\, \left( 3{\pi }^{2}-32
 \right) R^4{B_s}^{2}}{ \left( {\pi }^{2}-6 \right)^2 G{\rho_c}\,M^2}}
\ee
Scaling to canonical neutron star values for the different parameters, and introducing the volume averaged field as in the 
constant density case (here $\bar{B}=0.54 B_s$), we have 
\be
\epsilon\approx 8\times 10^{-11}\,\left(\frac{R}{10\mbox{ km}}\right)^4\,\left(\frac{M}{1.4 M_{\odot}}\right)^{-2}\,\left(\frac{\bar{B}}{10^{12}G}\right)^2
\label{epspol}\ee
In other words, for this model configuration the magnetic field induced ellipticity is almost two order of magnitude greater 
than in our constant density model. This is perhaps a useful indication of how much the result can vary for supposedly 
``similar'' field configurations if we change the equation of state. Of course, in making this statement one must keep in mind that the 
two field configurations really are different.

%%%%%%%%%%%%%%%%%%%%%%%%%%%%%%%%%%%%%%%%%%%%%%%%%%%%%%%%%%%%%
\subsection{n=1 polytrope with a toroidal field}
%%%%%%%%%%%%%%%%%%%%%%%%%%%%%%%%%%%%%%%%%%%%%%%%%%%%%%%%%%%%%%%%%%

We next consider the case of a purely toroidal magnetic field in an star with an $n=1$ polytropic equation of state.
For the field given in (\ref{toro1}) the Lorentz force has  $l=2$ components,
\beq
L_r&=&\frac{2}{3}\frac{{B}^{2}}{ {r}{\pi }^{2}{R}} \left[ R\,\sin \left( {\frac {r\pi }{R}} \right) +\pi \,r
\cos \left( {\frac {r\pi }{R}} \right)  \right] \sin \left( {\frac {r
\pi }{R}} \right)P_2(\theta)\nn\\
&&\nn\\
L_\theta&=&\frac{2}{3}\frac{{B}^{2}} { {r}{\pi }^{2}}\sin^2 \left( {\frac {r\pi }{R}}
 \right)\frac{dP_2}{d\theta}
\eeq
From equation (\ref{hydro2}) we can then write
\be
\delta\rho=-\frac{1}{24 {R}^{3}
{{\rho_c}}{G}{\pi }}\left[ 6\,{\rho_c}\,{\delta\Phi} \left( r \right) R\,{\pi }^{2}-{B}^{2}\sin \left( {\frac {r\pi }{R}} \right) r \right]
\ee
which, inserted on the right hand side of Poisson's equation (\ref{pertp}) gives\be
\delta\Phi=-\frac{1}{36}\,\left(\frac{B}{R}\right)^{2} \left[ \pi\,(r^5\pi^2-15rR^4)\cos \left( {\frac {\pi \,r}{R}} \right) \,+5R^3\,(R^2-r^2\pi)\,\sin \left( {\frac {\pi \,r}{R}} \right) \right] \frac{1}{{\pi }^{4}{{\rho_c}}{r}^{3}}
\ee
We can then calculate the distortion at the surface
\be
\varepsilon=-{\frac {1}{144}}\,{\frac {{B}^{2} \left( {\pi}^2-15 \right) }{{\pi }^{2}{R}^{2}G{{\rho_c}}^{2}}}
\ee

It follows that the  ellipticity
 is given by
\be
\epsilon={\frac {1}{9\pi}}\,\frac{(\pi^2-15)}{(\pi^2-6)}{\frac {R^4\,{B}^{2}
}{G{\rho_c}\,M^2}}
\ee
or using the average field, which in this case corresponds to $\bar{B}\approx-0.17 B$,
\be
\epsilon\approx-5\times 10^{-12}\left(\frac{R}{10\mbox{ km}}\right)^4\,\left(\frac{M}{1.4\,M_{\odot}}\right)^{-2}\,\left(\frac{\bar{B}}{10^{12}G}\right)^2 
\label{epstor}\ee
The deformation is now prolate, and notably about one order of magnitude smaller than in the case of a poloidal field (for the same $\bar{B}$).
We can compare this result to the estimate used by \citet{cutler2002}. 
For a $10^{15}$~G toroidal field he assumes that 
$\epsilon\approx - 1.6 \times 10^{-6}$ (the same magnitude as in the poloidal case). The deformation we have determined is about a factor of three larger.

\subsection{General deformations}

Having considered some particular examples, let us now present the formalism for the case of a more 
general field configuration and equation of state. Assume that the
 magnetic field takes the form of equation (\ref{genfield}), for which the Lorentz force is
\beq
L_r&=&-\frac{dA}{dr}\frac{\sin^2\theta}{r^4}\left[(\lambda^2r^2-2)A+\frac{d^2A}{dr^2}\right]\nn\\
L_\theta&=&=-\frac{2A}{r^5}\cos\theta\sin\theta\left[(\lambda^2r^2-2)+\frac{d^2A}{dr^2}\right]
\eeq
Focusing our attention on the $l=2$ components, we have (see Appendix~B for a discussion of the associated radial deformation)
\beq
L_r(l=2)&=&\frac{dA}{dr}\frac{2P_2(\theta)}{3r^4}\left(\lambda^2Ar^2+\frac{d^2A}{dr^2}-2A\right)\nn\\
L_\theta(l=2)&=&=\frac{2A}{3r^5}\frac{dP_2}{d\theta}\left(\lambda^2Ar^2+\frac{d^2A}{dr^2}-2A\right)
\label{elle2}
\eeq
From the angular part of the perturbed hydrostatic equilibrium equations (\ref{hydro1}) we obtain (for the $l=2$ case we are considering)
\beq
\delta p&=&-\rho\delta\Phi+rL_\theta\nn\\
&=&-\rho\delta\Phi+\frac{2}{3}\frac{A}{r^4}\left(\lambda^2Ar^2+\frac{d^2A}{dr^2}-2A\right)
\eeq
which substituted back into the radial part of the equations leads to
\beq
\delta\rho&=&-\frac{1}{3r^5\frac{d\Phi}{dr}}\left[-3\frac{d\rho}{dr}\delta\Phi r^5+r\frac{dA}{dr}A\,(2r^2\lambda^2-4)+4A^2\,(4-r^2\lambda^2)+2r^3A\frac{d^3A}{dr^3}-4r^2A\frac{d^2A}{dr^2}\right]
\label{diro2}\eeq
This now allows us to compute the source term of the perturbed Poisson equation (\ref{pertp}), which, for $l=2$, reads
\beq
&&\frac{d^2\delta\Phi}{dr^2}+\frac{2}{r}\frac{d\delta\Phi}{dr}-\delta\Phi\left[\frac{6}{r^2}+\frac{4\pi G}{r^5\frac{d\Phi}{dr}}\!\left(\!\frac{d\rho}{dr}r^5\right)\right]=\!-\frac{4\pi G}{3r^5\frac{d\Phi}{dr}}\!\left[r\frac{dA}{dr}A(2r^2\lambda^2-4)+4A^2(4-r^2\lambda^2)+2r^3A\frac{d^3A}{dr^3}-4r^2A\frac{d^2A}{dr^2}\right]
\eeq
The boundary conditions for this equation remain regularity at the centre and continuity of $\delta\Phi$ and its derivative at the surface. 
For $l=2$ the exterior gravitational potential has the form
\be
\delta\Phi_\mathrm{ext}\propto\frac{1}{r^3}
\label{epsi1}\ee
Having computed $\delta\Phi$ in the interior we can obtain $\delta\rho$ from equation (\ref{diro2}) and thus compute the ellipticity $\epsilon$ and the deformation at the surface
\be
\varepsilon(R)=-\frac{\delta\rho(R)}{R\frac{d\rho}{dr}(R)}
\label{epsi2}\ee

As an example we carry out the calculation for a polytrope and a mixed poloidal/toroidal field. Then we can use the 
stream function from equation (\ref{tutto1}) to determine the magnetic field and the Lorentz force.
 The result will obviously depend on which of the eigenvalues we choose for $\lambda$. 
This way we can vary the relative strength of the toroidal part of the field.
A sample of the obtained results is listed in Table~\ref{polotoro}. As expected, for low values of $\lambda$ the effects of the poloidal and toroidal field work against each other and the deformation is comparatively small. It is possibly surprising that the cancellation is so 
efficient. If one were to, for example, add the deformations deduced for a pure poloidal field (\ref{epspol}) to that for a purely toroidal field 
(\ref{epstor}) using the appropriate values for $\bar{B}$ then one would predict that the deformation should be $\epsilon\sim 10^{-11}$ 
for the first eigenvalue in the Table. Furthermore, one would have expected the deformation to be prolate. We see that for the mixed field we are using 
the true deformation is almost two orders of magnitude smaller. It is also toroidal. This illustrates that it will be important to 
use both realistic field configurations and equations of state in estimated magnetic neutron star deformations.
The results in Table~\ref{polotoro} show that the star is always prolate, and as $\lambda$ grows the toroidal field becomes dominant and the star becomes more and more prolate. Once a toroidal component is introduced, the poloidal component of the field is, thus, never strong enough to make the star oblate
for this model.

\begin{table}
\begin{tabular}{l|l|l}
$\lambda$&$\epsilon$&$\bar{B}_t$\\
\hline
7.420&$-6.3\times 10^{-13}$&$3.0\times 10^{12}$\\

10.706&$-2.8\times 10^{-12}$&$4.2\times 10^{12}$\\

23.433&$-4.4\times 10^{-11}$&$8.1\times 10^{12}$\\

80.073&$-2.3\times 10^{-10}$&$2.7\times 10^{13}$\\

105.215&$-4.4\times 10^{-10}$&$3.5\times 10^{13}$\\

183.994&$-3.7\times 10^{-8}$&$9.3\times 10^{13}$\\

1000.59&$-1.3\times 10^{-6}$&$4.8\times 10^{14}$
\end{tabular}
\caption{The ellipticity $\epsilon$ for a sample of eigenvalues  $\lambda$. The results correspond to the parameters $M=1.4M_{\odot}$, $R=10$~km and we take the energy averaged amplitude of the poloidal part of the field to be $10^{12}$~G. The energy averaged toroidal field is
 given in each case. As can be seen, the star starts off with a small deformation as the effects of the poloidal and toroidal components cancel each other, and becomes more and more prolate as the toroidal component of the field grows. Note that we have a vanishing exterior field in all cases.}
\label{polotoro}
\end{table}

\subsection{n=1 polytrope, field confined to the crust}

Finally, let us consider the case when the field is confined to a region close to the surface.
As we have already discussed, 
we shall call this ``crust'', even though we are neglecting its elastic properties.
We consider the core to be unperturbed, and impose the condition (\ref{crust2}) at the inner boundary.
Then the deformation is confined to the crust and, as the density is low in this region, the quadrupole moment, and thus the ellipticity, will in general be much smaller than if we were considering the deformations of the whole star. However, it is possible that the field is confined to the crust after being expelled from a superconducting core. In this case a large number of field lines are squeezed into a small region close to the surface of the star and would produce a strong field and large deformations. This result can be seen in table \ref{crustdef}, where we have calculated the deformation of the star by assuming that the total magnetic energy in the crust is equal to that of the field extended to the whole star, as calculated in the previous section. As can be seen the ellipticity is larger than in the case of a field extending throughout the star, by up to more than an order of magnitude. This agrees with what was found in \cite{BG1996}, where the authors analysed the deformations due to a poloidal field confined to a thin shell close to the surface of a neutron star. This motivates a more detailed investigation into the problem of superconducting neutron star cores.

\begin{table}
\begin{tabular}{l|l|l|l}
$\lambda$&$\epsilon$&$\bar{B}_p$&$\bar{B}_t$\\
\hline
103.60&$-1.0\times 10^{-7}$&$6.2\times 10^{14}$&$1.2\times 10^{15}$\\

496.39&$-5.5\times 10^{-7}$&$5.8\times 10^{14}$&$1.3\times 10^{15}$\\

1008.46&$-1.7\times 10^{-5}$&$5.6\times 10^{14}$&$1.4\times 10^{15}$\\

\end{tabular}
\caption{The ellipticity $\epsilon$ for various eigenvalues $\lambda$. The results correspond to the parameters $M=1.4M_{\odot}$, $R=10$~km. We take the field to have the same energy as that of a field extended to the whole star with $\lambda=1000.59$. The energy averaged values of the poloidal and toroidal magnetic field are also given. We should thus compare all results for the ellipticity in this table with the one given by $\lambda=1000.59$ in Table~\ref{polotoro}. The field is confined to a region close to the surface beginning at a radius $r_b=9\times 10^5$ cm, thus roughly corresponding to a 1 km thick crust. The core is taken to be spherical. As can be seen for the listed results, the deformations are larger than in the case of the field extending to the whole star, by up to an order of magnitude.}
\label{crustdef}
\end{table}

\section{Concluding discussion}

We have presented a scheme for calculating the magnetic deformations of a neutron star. We have considered the particular case of a dipolar field in a non-rotating star, but the formalism is readily extended to other cases. The formalism can, for example, be adapted to the case of slow rotation if we take the magnetic axis to be aligned with the spin axis of the star (see Appendix~A). We have considered the case of a purely poloidal field, the effect of which is to make the star oblate, that of a purely toroidal field which makes the star prolate, and the case of a mixed poloidal and toroidal field, which makes the star more and more prolate as the strength of the toroidal component is increased. We have seen that  the condition that the 
magnetic field be consistent with the background equation of state is very restrictive. For the present set of model problems,  we are forced to consider only fields that vanish outside the star. This is clearly not physical as one would expect the field to be prevalently dipolar far from the star. This problem could be solved by introducing surface currents. Since we do not have a physical model for such currents we have not considered 
this possibility in detail. Nevertheless, 
our formalism is sufficiently general that it could be extended to any field configuration.

It  seems appropriate to conclude this investigation by discussing the relevance of our results for gravitational-wave observations.
A rotating magnetic neutron star  becomes interesting from the gravitational-wave point of view when the magnetic axis is  
not aligned with the spin axis. Then the deformation is no longer be axisymmetric and we have a time varying quadrupole. 
This is in contrast with the typically much larger (see Appendix~A) rotational deformation which is always axisymmetric 
and cannot radiate gravitational waves. 
The framework we have described can, in principle, be used with any equation of state or field configuration for the neutron star. 
It allows us to calculate the ellipticity, and thus the quadrupole, in order to make quantitative gravitational-wave estimates.
This is a useful exercise since we can compare our results to observational upper limits on the neutron star ellipticity, and perhaps
constrain the parameters of proposed theoretical models. To see how this works out in the present case, let us 
consider the recent upper limits on pulsar signals from the LIGO effort \citep{ligo}.   The strongest current constraint comes from 
PSR J2124-3358, a millisecond pulsar spinning at $202.8$~Hz. For this system, LIGO obtains the constraint $\epsilon<8.5\times 10^{-7}$. 
This is impressive, because it restricts the maximal neutron star mountain to a fraction of a centimetre. The question is, how much 
does this constrain the magnetic field? From (\ref{epspol}) and (\ref{epstor}) we easily find that we must have
\be
\bar{B} <  \left\{ \begin{array}{ll} 
10^{14} \mbox{ G   for a purely poloidal field} \\
4\times10^{14} \mbox{ G   for a purely toroidal field} 
\end{array}\right.
\label{constraints}\ee
(taking radius and mass to have the canonical values). Is this constraint telling us anything interesting?  The answer is probably no. 
The pulsars currently considered by LIGO are almost exclusively 
fast spinning, taken from the millisecond spin period sample. For these 
neutron stars the exterior dipole field inferred from the spindown is weak. In the case of PSR J2124-3358 the 
standard estimate would suggest that the exterior field is about $3\times 10^8$~G. To understand the relevance of this we must first 
recall that our model stars have a vanishing exterior field. Thus there is something missing in the model. However, 
it is thought that the interior field could be significantly stronger than that in the exterior.   
It could well be that our simple models  provide  useful representations of this interior field, and what is missing is a relatively 
small part that is irrelevant when it comes to deforming the star. Nevertheless, it does not seem reasonable to suggest that the 
interior field can be as much as six orders of magnitude stronger than the exterior one (for reasons of stability etcetera). 
Hence, our results indicate that LIGO has not yet reached the 
sensitivity where it constrains any reasonable theory. Given this, one may ask to what extent future observations will improve on the current 
results. It is relatively straightforward to estimate how well one can hope do to if we recall that (using matched filtering) the
effective gravitational-wave amplitude increases as the square-root of the number of observed wave cycles. For a spinning neutron star,
where the spin frequency is essentially fixed during the observation, this means that the signal-to-noise ratio 
scales with the observation time as 
$t_\mathrm{obs}^{1/2}$. The data used by \cite{ligo} is based on observations lasting the order of three weeks. If one could extend this
to one full year, one would gain a factor of 4 or so in amplitude. This translates into an improvement of a factor 
of 2 in the constraint on the magnetic field,  not enough to test a realistic theory model. 
In order to do much better one would have to improve the detector sensitivity. With the advanced LIGO upgrade, the 
sensitivity will increase by about one order of magnitude. With a one year observation we then get a further $\sqrt{10}$ improvement 
in the constraint on the magnetic field. Advanced detectors will also have the ability to narrow-band and focus on a particular 
frequency. Suppose one were to do that and get (say) an additional order of magnitude improvement of sensitivity at the particular 
frequency for a given pulsar. This would still only improve the constraints in (\ref{constraints}) by a factor of about 
$2\times\sqrt{10}\times\sqrt{10}=20$. 
These rough estimates suggest that we should probably not expect to detect gravitational waves from magnetically deformed
millisecond pulsars. There are obvious caveats to this, in particular, associated with the superconductivity of the core. If all the 
magnetic flux is confined to the crust then the deformation may be larger, cf. the results in Table~\ref{crustdef}. 
This problem is clearly worth further investigation. 

The situation is not quite as pessimistic if one considers slower spinning pulsars. Consider for example the current LIGO upper limit 
of $\epsilon < 2\times 10^{-3}$ for the Crab pulsar \citep{ligo}.
For the models we have considered we must then have
\be
\bar{B} <  \left\{ \begin{array}{ll} 
5\times 10^{15} \mbox{ G   for a purely poloidal field} \\
2\times10^{16} \mbox{ G   for a purely toroidal field} 
\end{array}\right.
\ee
At first sight, these limits may seem less interesting than those from the faster spinning system. However, we need to remember that 
young pulsars, like the Crab, are thought to have much stronger magnetic fields. From the spin-down one would usually estimate that 
the exterior dipole field is $4\times 10^{12}~G$ for the Crab. Hence, the current observed upper limit is 
only a factor of about a factor of 1000 away from the expected field. One should be able to improve this 
by at least an order of magnitude with future detectors (the current LIGO detectors may be improved at lower frequencies, Advanced LIGO
will do better and VIRGO is expected to have good performance in the low-frequency regime) and a full year of observation.
This suggests that gravitational-wave observations may in the future be able to test theoretical models where 
the interior magnetic field is more than two orders of magnitude stronger than the exterior field. 

The challenge now is to improve on the rough estimates we have presented in this paper. Future work needs 
to consider the deformation for realistic field
configurations such as those worked out by \cite{spruit}. We also need to understand the role of superconductivity better. 
There are a number of interesting questions associated with accreting systems. These involve the burial of the field by 
accreted matter, leading to a relatively weak exterior field, and potential accretion induced asymmetries \citep{melatos}. 
Finally, it may also be interesting to consider magnetars. After all, they are expected to have fields as strong as the contraints 
given in (\ref{constraints}). Of course, they are also spinning very slowly which places them outside the bandwidth of any groundbased 
gravitational-wave detector. It is still possible that a newly born magnetar may occasionally pass through the detection window.
Such an event should be observable from within the galaxy, and perhaps beyond.

\section*{Acknowledgments}

This work was supported by PPARC through grant numbers PPA/G/S/2002/00038 and
PP/E001025/1.
BH would like to thank the Institut Henri Poincare-Centre Emile Borel for hospitality and support during part of this work.
NA  acknowledges support from PPARC via Senior Research Fellowship no
PP/C505791/1. LS gratefully acknowledges support from a Mare Curie Intra-European Fellowship, contract number MEIF-CT-2005-009366.

\appendix
\section{General deformations: magnetic fields and rotation}

The  framework discussed in the main body of the paper
can easily be extended to the case where the deformation is due not only to a magnetic field, but also to the star's rotation.
We can write then the equations of motion as
\be
\frac{\boldsymbol{\nabla} p}{\rho}+\boldsymbol{\nabla}{\psi}=\frac{(\boldsymbol{\nabla}\times \mathbf{B})\times \mathbf{B}}{4\pi\rho}
= { \mathbf{L} \over 4 \pi \rho}
\ee
where
\be
\psi=\Phi-1/2\Omega^2r^2\sin^2\theta
\ee
$\Phi$ is the gravitational potential and $\Omega$ the constant rotation rate.
As the new term due to rotation is written as the gradient of a scalar function, it's curl will still vanish and the compatibility condition for the magnetic field in equation (\ref{field}) remains the same. We can thus still use the magnetic field from equation (\ref{genfield}).
The equations of hydrostatic equilibrium, for the $l=2$ case, now take the form
\beq
&&\frac{d\delta p}{dr}+\rho\frac{d\delta\Phi}{dr}+\delta\rho\frac{d\Phi}{dr}+\frac{2}{3}\rho\Omega^2r=L_r\nn\\
&&\delta p+\rho\delta\Phi+\frac{\rho\Omega^2 r}{3}=rL_\theta
\label{tutto}\eeq
where $L_r$ and $L_\theta$ are given in equation (\ref{elle2}).
We can proceed as in the previous analysis and use equations (\ref{tutto}) to obtain $\delta\rho$ as a function of $\delta\Phi$, $B^2$ and $\Omega$.
This allows us to write the perturbed Poisson equation (for $l=2$) as:
\beq
&&\frac{d^2\delta\Phi}{dr^2}+\frac{2}{r}\frac{d\delta\Phi}{dr}-\delta\Phi\left[\frac{6}{r^2}+\frac{4\pi G}{r^5\frac{d\Phi}{dr}}\!\left(\!\frac{d\rho}{dr}r^5\right)\right]=\!-\frac{4\pi G}{3r^5\frac{d\Phi}{dr}}\!\left[r\frac{dA}{dr}A(2r^2\lambda^2-4)+4A^2(4-r^2\lambda^2)+2r^3A\frac{d^3A}{dr^3}\right.\nn\\
&&\left.-4r^2A\frac{d^2A}{dr^2}+\frac{d\rho}{dr}\Omega^2r^7\right]
\eeq
and thus obtain the surface deformation $\varepsilon$ and the ellipticity $\epsilon$ as defined in equations (\ref{epsi1}) and (\ref{epsi2}).

As an example let us take an $n=1$ polytrope with the purely poloidal field from equation (\ref{n1}). The equations of hydrostatic equilibrium now include the rotation term as in equation (\ref{tutto}), where the Lorentz force is that of equation (\ref{lorentzn1}).
We can thus apply exactly the same procedure as previously and solve for the perturbed quantities $\delta\Phi$, $\delta\rho$ and $\delta p$.
This allows us to calculate the deformation at the surface
\be
\varepsilon=-\frac{\pi^5B^2(24-\pi^2)}{16R^2\rho_c^2G(\pi^2-6)^2}-\frac{5}{4\pi}\frac{\Omega^2}{G\rho_c}
\ee
Working out the corresponding ellipticity we find
\be
\epsilon=8\times 10^{-11}\left(\frac{R}{10 \mbox{ km}}\right)^2\,\left(\frac{M}{1.4\,M_\odot}\right)^{-4}\,\left(\frac{\bar{B}}{10^{12}G}\right)^2+ 1.4\times 10^{-1}\left(\frac{\Omega}{\Omega_{br}}\right)^2
\ee
where $\Omega_{br}=\frac{2}{3}\sqrt{\pi G\bar{\rho}}$ is the breakup frequency ($\bar{\rho}$ is the average density of the non-rotating model). For our values this frequency would be $f\approx$ 1250 Hz, which corresponds to a period $P\approx 0.8$ ms.
We see that, as expected, the rotational term completely dominates the magnetic term. Rotational deformations will, however, always be axisymmetric. The magnetic deformations can  lead to a quadrupolar deformation and thus to gravitational wave emission only if the magnetic axis is inclined with respect to the spin axis.
 
\section{Radial deformations}

As discussed in Section~5, it may be relevant to work out also the radial deformation due to the presence of the magnetic field. 
For general fluid configurations, any quadrupole deformation is likely to be accompanied by an $l=0$ component. 
In order to determine the radial deformation we can assume that the new radial variable $x$ defined in equation (\ref{newvar}) labels the deformed gravitational equipotential surfaces. This mean that we impose that
\be
\nabla x\times \nabla\Phi=0
\ee
which, if we write
\be
\varepsilon(x,\theta)=D_0(x)+D_2(x)P_2(\theta)
\ee
 leads to the condition
\be
D_2=-\sqrt{\frac{4\pi}{5}}\frac{\delta\Phi}{r\frac{d\Phi}{dr}}\label{d2}
\ee
This expression allows us to calculate the quantity $D_2$ throughout the star and can be of use, for example, when discussing oscillations on a deformed background,
cf. \cite{saio}.
Writing out the equations that need to be solved in the $l=0$ case, the perturbed Euler equation and Poisson equation, we have
\beq
\frac{d}{dr}\delta\Phi_0&=&\psi_0\nn\label{defo}\\
\frac{d}{dr}\delta\psi_0&=&4\pi G \delta\rho_0-\frac{2}{r}\psi_0\nn\\
\frac{d}{dr}\delta\rho_0&=&\frac{\rho}{P}\left[L_0(B)-\rho\psi_0-\left(\frac{d}{dr}\left(\frac{P}{\rho}\right)\Gamma_1+\frac{d}{dr}\Phi\right)\right]
\eeq
where $\psi_0$ is introduced in order to obtain a first order system. We have also assumed a  linearised barotropic equation of state 
\be
\frac{\delta P}{P}=\Gamma_1\frac{\delta\rho}{\rho}
\ee
and $L_0(B)$ is the $l=0$ part of the Lorentz force, which can be calculated once the magnetic field has been specified.
These equations can now be integrated, imposing regularity at the centre of the star and imposing that at the surface $\delta\rho_0=0$ and matching $\delta\Phi_0$ and $\delta\Phi_0'$ to an exterior solution of the form $\delta\Phi_0^\mathrm{ext}\propto 1/r$.
This will allow us to calculate
\be
D_0=-\sqrt{{4\pi}}\frac{\delta\Phi_0}{r\frac{d\Phi}{dr}}
\label{d0}
\ee

\end{document}